\documentclass[reprint,
superscriptaddress,
onecolumn,
showkeys,
 amsmath,amssymb,
 aps,
 prf,
]{revtex4-2}
\usepackage{graphicx}
\usepackage{dcolumn}
\usepackage{bm}
\usepackage[dvipsnames]{xcolor}
\usepackage{caption}
\usepackage{subcaption}
\begin{document}
\title{Capillary-lubrication force exerted on a 2D particle moving towards a fluid interface}
\author{Aditya Jha}
\affiliation{Univ. Bordeaux, CNRS, LOMA, UMR 5798, F-33400 Talence, France.}
\author{Yacine Amarouchene}
\affiliation{Univ. Bordeaux, CNRS, LOMA, UMR 5798, F-33400 Talence, France.}
\author{Thomas Salez}
 \email{thomas.salez@cnrs.fr}
\affiliation{Univ. Bordeaux, CNRS, LOMA, UMR 5798, F-33400 Talence, France.}
\date{\today}
\begin{abstract}
A rigid object moving in a viscous fluid and in close proximity with an elastic wall experiences self-generated elastohydrodynamic interactions. This has been the subject of an intense research activity, with a recent and growing attention given to the particular case of elastomeric and gel-like substrates. Here, we address the situation where the elastic wall is replaced by a capillary surface. Specifically, we analyze the lubrication flow generated by the prescribed normal motion of a rigid infinite cylinder near the deformable interface separating two immiscible and incompressible viscous fluids. Using a combination of analytical and numerical treatments, we compute the emergent capillary-lubrication force at leading order in compliance, and characterize its dependencies with the interfacial tension, viscosities of the fluids, and length scales of the problem. Interestingly, we identify two main contributions: i) a velocity-dependent adhesion-like force ; ii) an acceleration-dependant inertial-like force. Our results may have implications for the mobility of colloids near complex interfaces and for the motility of confined microbiological entities.
\end{abstract}
\maketitle

\section{\label{sec:intro}Introduction}
The motion of a rigid object in a fluid has been well studied during the last couple of centuries~\cite{Batchelor1967}. In a bulk situation, for an incompressible Newtonian viscous fluid, the hydrodynamic force exerted on the object depends on its shape, size and speed, as well as on the fluid viscosity and frictional boundary conditions. Adding a neighbouring rigid wall to the latter problem was an obvious extension to consider, in view of the historical importance of lubricated contact mechanics in industry, but also because of the modern trends in miniaturization, colloidal surface science and confined biological physics. Such a modification introduces a symmetry breaking as well as different flow boundary conditions~\cite{o1967slow,goldman1967slow, cooley1969slow,jeffrey1981slow}. Happel and Brenner provided a detailed account of this situation, including the related case of suspensions~\cite{happel1983low}. 

In view of the growing interest in soft matter towards complex materials, such as elastomers, gels, or biological membranes, replacing the above rigid wall by an elastic boundary became of central importance. In such a context, the influence of the elastic response on the lubrication flow and associated forces and torques
was addressed in both the normal~\cite{Balmforth2010,leroy2011hydrodynamic,leroy2012hydrodynamic,Villey2013,wang2015out,Karan2018} and transverse modes~\cite{sekimoto1993mechanism,Beaucourt2004,skotheim2005soft,Weekley2006,urzay2007elastohydrodynamic,Snoeijer2013,salez2015elastohydrodynamics,Bouchet2015,Saintyves2016,Davies2018,Rallabandi2018,Vialar2019,zhang2020direct,Essink2021,bertin2022soft,Bureau2023}. This was essentially achieved by combining previous works on: i) solid-solid contact and linear elasticity ~\cite{Johnson1985,li1997elastic,nogi1997influence,nogi2002influence}, and ii) lubrication theory~\cite{Reynolds1886,
Oron1997}, resulting in the so-called soft-lubrication theory. These developments led in part to the design of non-invasive contactless mechanical probes for the rheology of soft, fragile and alive materials~\cite{garcia2016micro, Basoli2018}. Further studies then incorporated elements of complexity in the substrate's response, through \textit{e.g.} viscoelasticty~\cite{pandey2016lubrication,guan2017noncontact,Kargar2021, zhang2022contactless} and poroelasticity~\cite{kopecz-muller2023}. 

Interestingly, as materials get softer and increasingly liquid-like, solid capillarity takes the relay over bulk elasticity to eventually become the dominant restoring mechanism -- a topic of recent and active research~\cite{Andreotti2016}. As a consequence, investigating soft-lubrication-like couplings in situations where the flow-induced interfacial deformation is mainly resisted by surface tension appears to be a relevant task. In a series of seminal articles, Lee, Leal and colleagues calculated the forces felt by a sphere moving close to a fluid interface in Stokes flow~\cite{lee1979motion,lee1980motion,berdan1982motion,lee1982motion,geller1986creeping}. Using Lorentz's reciprocal theorem, as well as a complete eigenfunction expansion in bipolar coordinates, they were able to exhibit the effects of the fluid interface -- albeit in the regime where the gap between the sphere and the interface is large and the interfacial deformation is negligible. Related developements included the cases of slender objects~\cite{yang1983particle}, bubbles and droplets~\cite{Vakarelski2010,chan2011film}, living microorganisms~\cite{trouilloud2008soft,lopez2014dynamics}, slippery interfaces~\cite{Rinehart2020}, and air-water interfaces with surface-active contaminants~\cite{Maali2017,bertin2021contactless}.

While the above studies clearly highlight the richness and importance of motion near fluid interfaces, they focus on specific geometries and viscosity ratios. Hence, the general capillary-lubrication regime has only been scarcely explored so far. In the present article, we thus theoretically and numerically investigate the lubrication flow and associated force generated by the prescribed normal motion of a rigid infinite cylinder near a deformable interface separating two immiscible and incompressible viscous fluids. We invoke a perturbative approach in dimensionless compliance, and study the influence of the interfacial tension, viscosities of the fluids, and length scales of the problem on the resulting capillary-lubrication force. 

The article is organized as follows. We first start by setting the general capillary-lubrication theoretical framework. Then, the perturbation analysis is presented for the pressure and deformation fields up to first order in dimensionless compliance, the latter being directly related to the capillary number. Finally, we discuss the results, compute quantitatively the capillary-lubrication force and investigate the influence of all physical and geometrical parameters on the latter.

\section{Capillary-lubrication theory}
\begin{figure}[h]
\begin{center}
\includegraphics[width=8cm]{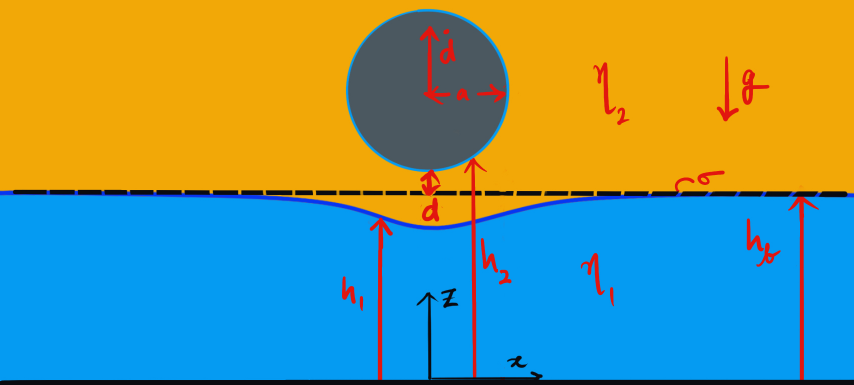}
\caption{Schematic of the system. A rigid infinite cylinder moves with a prescribed velocity normally to a nearby capillary interface between two incompressible Newtonian viscous liquids. The ensemble is placed atop a rigid substrate. The origin of spatial coordinates is located at the interface between the rigid substrate and the bottom liquid layer ($z=0$) under the center of mass of the cylinder ($x=0$).}
\label{fig:1}
\end{center}
\end{figure}

We consider a rigid infinite cylinder of radius $a$ moving with a prescribed but potentially time-dependent velocity normally to a nearby fluid interface, as shown in Fig.~\ref{fig:1}. The interface is characterized by its surface tension $\sigma$, and separates two incompressible Newtonian viscous liquids, with dynamic shear viscosities $\eta_1$ and $\eta_2$, as well as densities $\rho_1$ and $\rho_2=\rho_1-\delta\rho$ (with $\delta\rho>0$). The acceleration of gravity is denoted $g$. The thickness profile $h_1(x,t)$ of the bottom liquid layer depends on the horizontal position $x$ as well as time $t$, and at large $x$ it equals the undeformed reference value $h_\textrm{b}$. The total thickness profile between the rigid substrate and the cylinder surface is denoted by $h_2(x,t)$. We also define the minimal distance $d(t)=h_2(0,t)-h_\textrm{b}$ between the undeformed fluid interface and the cylinder surface, the time derivative $\dot{d}(t)$ of which being the prescribed time dependent velocity of the cylinder along $z$.

\subsection{Governing equations}
We neglect fluid inertia and assume the typical thicknesses, \textit{e.g.} $h_1(0,t)$ and $h_2(0,t)-h_1(0,t)$, of the two relevant liquid films of the problem to be much smaller than the proper horizontal length scale -- whether the latter is the cylinder radius $a$, the capillary length $\sqrt{\sigma/(g\delta\rho)}$, or the hydrodynamic radius $\sqrt{2ad}$~\cite{leroy2011hydrodynamic}, as discussed below. Therefore, we can invoke the lubrication theory~\cite{Reynolds1886,
Oron1997}. Introducing the excess pressure fields $p_i(x,z,t)$ with respect to the hydrostatic contributions, and the horizontal velocity fields $u_i(x,z,t)$, in the two liquids indexed by $i=1,2$, the incompressible Stokes equations thus read at leading lubrication order:
\begin{align}
\frac{\partial p_i}{\partial z} &= 0,\label{eq:1}\\
\frac{\partial p_i}{\partial x} &= \eta_i\frac{\partial^2 u_i}{\partial z^2}.\label{eq:2}
\end{align}
Besides, since the dominant flow is typically located only in the lubricated-contact region underneath the cylinder, we approximate the shape of the cylindrical surface by its
parabolic expansion, leading to:
\begin{align} 
h_2(x,t) \simeq h_\textrm{b}+d(t)+\frac{x^{2}}{2 a}.\label{eq:8}
\end{align}.

Finally, we close the set of equations by setting the flow boundary conditions. We impose no slip at the three interfaces,
as well as tangential and normal stress balances at the fluid interface. Hence, at $z = 0$:
\begin{align}
u_{1} = 0, \label{eq:7}
\end{align}
while, at $z = h_1$, at leading lubrication order and under a small-slope assumption, one has: 
\begin{align}
u_{2} & = u_1,\label{eq:5}\\
\eta_2\frac{\partial u_2}{\partial z} & = \eta_1\frac{\partial u_1}{\partial z}, \label{eq:6}\\
p_2-p_{1}& \simeq\sigma \frac{\partial^2 h_1}{\partial x^2}+g(h_{\textrm{b}}-h_1)\delta\rho,\label{eq:19}
\end{align}
and, at $z = h_2$:  
\begin{align}
u_{2} = 0. \label{eq:4}
\end{align}

Let us now non-dimensionalize the equations through: 
\begin{align*}
h_1(x,t)& = d^*H_1(X,T),   & h_2(x,t)& = d^*H_2(X,T),     & x & =lX,  &  z & =d^*Z, \\
t& = \frac{d^*}{c}T, & d(t)& =d^*D(T), & u_1(x,z,t)& =\frac{lc}{d^*}U_1(X,Z,T), &u_2(x,z,t)& =\frac{lc}{d^*}U_2(X,Z,T),\\
p_1(x,t) &=\frac{\eta_2c l^2}{d^{^*3}}P_1(X,T), & p_2(x,t)&=\frac{\eta_2cl^2}{d^{^*3}}P_2(X,T), & h_{\textrm{b}}& = d^*H_\textrm{b}, & \dot{d}(t) &=c\dot{D}(T),
\end{align*}
with the hydrodynamic radius $l=\sqrt{2ad^*}$, and where $d^*$ and $c$  represent some characteristic vertical length and vertical velocity scales, that can be set to \textit{e.g.} $d(0)$ and $\dot{d}(0)$, respectively. Moreover, the viscosity ratio is denoted by $M = \eta_1/\eta_2$. Using these dimensionless variables, Eq.~(\ref{eq:8}) becomes:
\begin{align} 
H_2(X,T)=H_\textrm{b}+D(T)+X^{2}.\label{eq:8nd}
\end{align}
Solving Eqs.~(\ref{eq:1},\ref{eq:2}) together with the boundary conditions of Eqs.~(\ref{eq:7}-\ref{eq:6},\ref{eq:4})
gives the velocity profiles: 
\begin{align}
U_{1}=-P_{2}^{\prime}\frac{\left(H_{2}-H_{1}\right)^{2} Z}{2\left[H_{1}+M\left(H_{2}-H_{1}\right)\right]}+P_{1}^{\prime}\left\{\frac{Z^{2}}{2 M}+\frac{Z}{\left[H_{1}+M\left(H_{2}-H_{1}\right)\right]}\left[H_{1}\left(H_{1}-H_{2}\right)-\frac{H_{1}^{2}}{2 M}\right]\right\},\label{eq:13}
\end{align}
\begin{align}
U_{2}=P_{2}^{\prime}\left[\frac{Z^{2}-H_{2}^{2}}{2}\right. \left.-\left(Z-H_{2}\right)\left\{H_{1}+\frac{M\left(H_{2}-H_{1}\right)^{2}}{2\left[H_{1}+M\left(H_{2}-H_{1}\right)\right]}\right\}\right]+P_{1}^{\prime}\frac{\left(Z-H_{2}\right) H_{1}^{2}}{2\left[H_{1}+M\left(H_{2}-H_{1}\right)\right]},\label{eq:14}
\end{align}
where the prime symbol corresponds to the partial derivative with respect to $X$. 
We then calculate the flow rates within the two liquid films, as:  
\begin{align} 
Q_{1} & =\int_{0}^{H_{1}} U_{1}\, \textrm{d} Z  =-P_{2}^{\prime}\frac{H_{1}^{2}\left(H_{1}-H_{2}\right)^{2}}{4\left[H_{1}+M\left(H_{2}-H_{1}\right)\right]}-P_{1}^{\prime}\frac{ H_{1}^{3}[H_{1}+4 M\left(H_{2}-H_{1}\right)]}{12 M[H_{1}+M\left(H_{2}-H_{1}\right)]},\label{eq:Q1}
\end{align}
\begin{align} 
Q_{2} & =\int_{H_{1}}^{H_{2}} U_{2}\, \textrm{d} Z  =-P_{2}^{\prime}\frac{\left(H_{2}-H_{1}\right)^{3}}{12}\frac{4 H_{1}+M\left(H_{2}-H_{1}\right)}{H_{1}+M\left(H_{2}-H_{1}\right)}-P_{1}^{\prime}\frac{H_{1}^{2}\left(H_{2}-H_{1}\right)^{2}}{4\left[H_{1}+M\left(H_{2}-H_{1}\right)\right]}.\label{eq:Q2}
\end{align}
Thanks to volume conservation, the flow rates allow us to write down the two thin-film equations, which read: 
\begin{align} 
\frac{\partial H_1}{\partial T}+Q_1^\prime = 0,\label{eq:thinfilm1}\\ 
\frac{\partial (H_2-H_1)}{\partial T}+Q_2^\prime = 0. \label{eq:thinfilm2}
\end{align}
Finally, Eq.~(\ref{eq:19}) reads in dimensionless form:
\begin{align}
H_1^{\prime\prime}+\textrm{Bo}(H_{\textrm{b}}-H_1)=\kappa\left(P_2-P_1\right), \label{eq:laplaceentire}
\end{align}
where $\textrm{Bo} = (l/l_{\textrm{c}})^2$ denotes the Bond number of the problem, with $l_{\textrm{c}}=\sqrt{\sigma/(g\delta\rho)}$ the capillary length, and where $\kappa= \textrm{Ca}/\epsilon^4$ 
is the dimensionless compliance of the fluid interface, with $\textrm{Ca} = \eta_2c/\sigma$ a capillary number and $\epsilon=d^*/l$ a small lubrication parameter. 

All together, since $H_2$ is known from Eq.~(\ref{eq:8nd}) and the prescribed $D(T)$, there are actually three unknown fields in the problem: $H_1$, $P_1$ and $P_2$. These obey the set of three coupled differential equations given by Eqs.~(\ref{eq:thinfilm1}-\ref{eq:laplaceentire}), together with the following symmetry and spatial boundary conditions: $P_{i}^{\prime} = 0$ and $H_1^{\prime} = 0$ at $X = 0$, as well as $P_{i} \rightarrow 0$ and $H_1 \rightarrow H_{\textrm{b}}$ at $X\rightarrow \infty$.
 
\section{Perturbation analysis}
Following the approach of previous soft-lubrication studies~\cite{sekimoto1993mechanism,skotheim2005soft, urzay2007elastohydrodynamic, salez2015elastohydrodynamics,pandey2016lubrication}, we assume that $\kappa\ll1$ and perform an expansion of the fields up to first order in $\kappa$, as: 
\begin{align}
    H_1 = H_{\textrm{b}} + \kappa \Delta+O(\kappa^2), \label{eq:21}\\
    P_1 =P_{10}+\kappa P_{11}+O(\kappa^2), \label{eq:22}\\
    P_2 =P_{20}+\kappa P_{21}+O(\kappa^2), 
    \label{eq:23}
\end{align}
where $\kappa\Delta$ is the deformation profile of the fluid interface at first order in $\kappa$, and where $\kappa^{j}P_{ij}$ is the excess pressure contribution of layer $i$ at perturbation order $j$. We further impose the following symmetry and spatial boundary conditions: $P_{ij}^{\prime} = 0$ and $\Delta^{\prime} = 0$ at $X = 0$, as well as $P_{ij} \rightarrow 0$ and $\Delta \rightarrow 0$ at $X\rightarrow \infty$.

\subsection{Zeroth-order solution}
At zeroth order in $\kappa$, the fluid interface is undeformed, and the bottom-film thickness profile is constant and equal to $H_{\textrm{b}}$. Equations~(\ref{eq:thinfilm1})~and~(\ref{eq:thinfilm2}) can then be solved analytically using the symmetry and boundary conditions on the pressure fields. This leads to: 
\begin{align}
    P_{10} = \frac{9M\dot{D}}{2H_{\textrm{b}}^2}\ln\left(1+\frac{1}{\xi}\right), \label{eq:26}\\
    P_{20} = \frac{9M^2\dot{D}}{2H_{\textrm{b}}^2}\left[\ln\left(1+\frac{1}{\xi}\right)-\frac{1}{\xi}-\frac{1}{6\xi^2}\right] , \label{eq:27}
\end{align}
where we have introduced the auxiliary variable $\xi(X,T) = M[D(T)+X^2]/H_{\textrm{b}}$. Interestingly, the $X$-dependencies of the zeroth-order excess pressure fields are logarithmic-like, which differs notably from the rigid-substrate case where the excess pressure reads $P_{\textrm{s}}=-3\dot{D}/(D+X^2)^2$~\cite{jeffrey1981slow}. 

\subsection{First-order solution}
At first order in $\kappa$, Eq.~(\ref{eq:laplaceentire}) reads: 
\begin{align}
     \Delta^{\prime\prime}-\textrm{Bo}\Delta = P_{20}-P_{10}  \label{eq:28}.
\end{align}
The formal solution of the latter equation, satisfying the above symmetry and boundary conditions, is:
\begin{align}
     \Delta =    \Delta_0\cosh\left(X\sqrt{\textrm{Bo}}\right)-\frac{1}{\sqrt{\textrm{Bo}}}\int_0^X\textrm{d}y\,\left[P_{20}(y)-P_{10}(y)\right]\sinh\left[(y-X)\sqrt{\textrm{Bo}}\right],
     \label{gensol}
\end{align}
where $\Delta_0=\Delta(X=0,T)=\left(1/\sqrt{\textrm{Bo}}\right)\int_0^{\infty}\textrm{d}y\,\left[P_{10}(y)-P_{20}(y)\right]\exp\left(-y\sqrt{\textrm{Bo}}\right)$ is the central deformation of the fluid interface. This solution can be numerically evaluated, for fixed parameters Bo, $M$, and $H_{\textrm{b}}$, and a prescribed $D(T)$ trajectory.

In order to rationalize the asymptotic behaviours of the numerical solution, and to evaluate the central deformation, we employ an asymptotic-matching method, which is a usual approach for capillary problems~\cite{james1974meniscus,lo1983meniscus,dupr2015shape}. To do so, we assume a scale separation between: i) an inner problem characterized by the horizontal length scale $l$, where the dominant lubrication flow is located, and where gravity is absent; and ii) an outer problem characterized by the capillary length $l_{\textrm{c}}$, where gravity regularizes the deformation. Specifically, we assume that $l\ll l_{\textrm{c}}$, \textit{i.e.} $\textrm{Bo}\ll1$. Let us first study the inner problem and associated inner solution $\Delta_{\textrm{in}}(X,T)$. Assuming that $X\sqrt{\textrm{Bo}}\ll1$, Eq.~(\ref{eq:28}) can be approximated by:
\begin{align}
     \Delta_{\textrm{in}}^{\prime\prime} = P_{20}-P_{10}.
\end{align}
The solution of this equation, satisfying the symmetry condition $\Delta_{\textrm{in}}'(X=0,T)=0$, reads:
\begin{multline}
     \Delta_{\textrm{in}} = \mathcal{A}+\frac{9 \dot{D}}{4 H_\textrm{b}^{2}}\left\{(M-1)MX^2 \ln\left[1+\frac{H_\textrm{b}}{M(D+X^2)}\right]+(1-M)(H_\textrm{b}+MD)\ln\left[H_\textrm{b}+M(D+X^2)\right]\right.\\
     +\left[H_{\textrm{b}}+(M-1) D\right]M \ln \left(D+X^{2}\right)   +4 \sqrt{D} M(1-M) X \tan ^{-1}\left(\frac{X}{\sqrt{D}}\right)   +4 M(M-1)X \sqrt{D+\frac{H_\textrm{b}}{M}}  \tan ^{-1}\left(\frac{X}{\sqrt{D+\frac{H_\textrm{b}}{M}}}\right)\\-\frac{2M H_\textrm{b} X}{\sqrt{D}} \tan ^{-1}\left(\frac{X}{\sqrt{D}}\right)-\frac{H_\textrm{b}^{2}}{6 D^{3 / 2}} X \tan ^{-1}\left.\left(\frac{X}{\sqrt{D}}\right)\right\}, \label{eq:innerdelta}
\end{multline}
where $\mathcal{A}$ is a function of $T$ only. The far-field behaviour of this inner solution reads: 
\begin{multline}
     \Delta_{\textrm{in}} \sim \mathcal{A}  +\frac{9 \dot D}{4 H_\textrm{b}^{2}} \left\{ 2H_\textrm{b} \ln (X)+\pi X \left[2 M(1-M) \sqrt{D} \right.\right.   \left.+2 M(M-1) \sqrt{D+\frac{H_\textrm{b}}{M}}-\frac{M H_\textrm{b}}{D^{1/2}}-\frac{H_\textrm{b}^{2}}{12D^{3 / 2}}     \right] \\ +  \left.\left[H_\textrm{b}(3-M)+(H_\textrm{b}+MD)(1-M)\ln(M)+\frac{H_\textrm{b}^{2}}{6 D} \right] \right\}. \label{eq:innerdeltalim}
\end{multline}

Let us now study the outer problem and associated outer solution $\Delta_{\textrm{out}}(X,T)$. For $X$ being large enough, Eq.~(\ref{eq:28}) can be approximated by: 
\begin{align}
     \Delta_{\textrm{out}}^{\prime\prime}-\textrm{Bo}\Delta_{\textrm{out}} = -\frac{9\dot{D}}{2H_{\textrm{b}}X^2}. \label{eq:29}  
\end{align}
We stress that it is essential here to keep a non-zero right-hand-side source term in the equation, in order to generate a logarithmic contribution as in the inner case. The solution of this equation, satisfying the boundary condition $\Delta_{\textrm{out}}\rightarrow0$ at $X\rightarrow\infty$, reads: 
\begin{align}
     \Delta_{\textrm{out}} =\mathcal{B}\textrm{e}^{-X\sqrt{\textrm{Bo}}}+\frac{9\dot{D}}{4H_{\textrm{b}}}\left(\textrm{e}^{-X\sqrt{\textrm{Bo}}}\int_{-\infty}^{X\sqrt{\textrm{Bo}} }\frac{\textrm{e}^t}{t}\,\textrm{d}t-\textrm{e}^{X\sqrt{\textrm{Bo}}}\int_{X\sqrt{\textrm{Bo}} }^{\infty}\frac{\textrm{e}^{-t}}{t}\,\textrm{d}t\right), \label{eq:outerdelta}  
\end{align}
where $\mathcal{B}$ is a function of $T$ only. The small-$X$ behaviour of this inner solution reads: 
\begin{align}
     \Delta_{\textrm{out}} \sim \mathcal{B}(1-X\sqrt{\textrm{Bo}})+\frac{9\dot{D}}{2H_{\textrm{b}}}\left[\gamma+\frac{1}{2}\ln(\textrm{Bo})+\ln(X)\right], \label{eq:outerdeltalim}
\end{align}
where $\gamma$ is the Euler constant. 

Matching Eqs.~(\ref{eq:outerdeltalim})~and~(\ref{eq:innerdeltalim}) allows us to determine the two unknown functions, as:  
\begin{align}
     \mathcal{A} = \mathcal{B}+\frac{9 \dot D}{4 H_{\textrm{b}}^{2}}\left[H_{\textrm{b}}(M-3)+(H_{\textrm{b}} +MD) (M-1)\ln(M)-\frac{H_{\textrm{b}}^{2}}{6 D} \right]+\frac{9 \dot D\gamma}{2 H_{\textrm{b}}}+\frac{9 \dot D}{4 H_{\textrm{b}}}\ln(\textrm{Bo}),\label{eq:37}
\end{align}
\begin{align}
     \mathcal{B} = \frac{9 \dot{D} \pi}{4 H_{\textrm{b}}^{2}\sqrt{\textrm{Bo}}} \left[2 M(M-1) \sqrt{D}   +2 M(1-M) \sqrt{D+\frac{H_{\textrm{b}}}{M}}+\frac{M H_{\textrm{b}}}{\sqrt{D}}+\frac{H_{\textrm{b}}^{2}}{12D^{3 / 2}}\right]. \label{eq:36}
\end{align}
In addition, using these matching conditions, the central deformation of the fluid interface can be evaluated from $\Delta_{\textrm{in}}(X=0,T)$ if $\textrm{Bo}\ll1$, as:
\begin{align}
\Delta_0=\mathcal{A}+\frac{9 \dot{D}}{4 H_\textrm{b}^{2}}\left\{(1-M)(H_\textrm{b}+MD)\ln\left(H_\textrm{b}+MD\right) +\left[H_{\textrm{b}}+(M-1) D\right]M \ln \left(D\right)\right\}.
     \end{align}

Finally, using Eqs.~(\ref{eq:26}),~(\ref{eq:27}),~and~(\ref{gensol}), as well as the above symmetry and boundary conditions, we can numerically solve Eqs.~(\ref{eq:thinfilm1})~and~(\ref{eq:thinfilm2}) at first order in $\kappa$, and hence compute the first-order pressure fields. The results are discussed below.

\section{Discussion}
Hereafter, keeping $\textrm{Bo}\ll1$, we discuss the zeroth-order and first-order solutions, and investigate the influence of the two key parameters: the viscosity ratio $M$, and the thickness ratio $H_{\textrm{b}}$. 

\subsection{Zeroth-order pressure}
\begin{figure}[h]
     \centering
     \begin{subfigure}{0.49\textwidth}
         \centering
         \includegraphics[width=7cm]{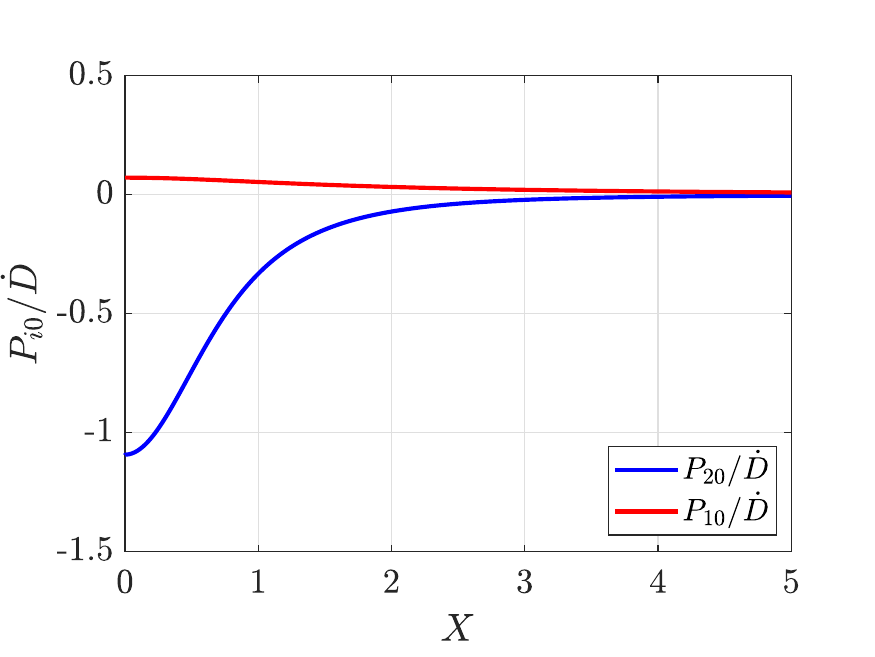}
         \caption{}
     \end{subfigure}
     \begin{subfigure}{0.49\textwidth}
         \centering
         \includegraphics[width=7cm]{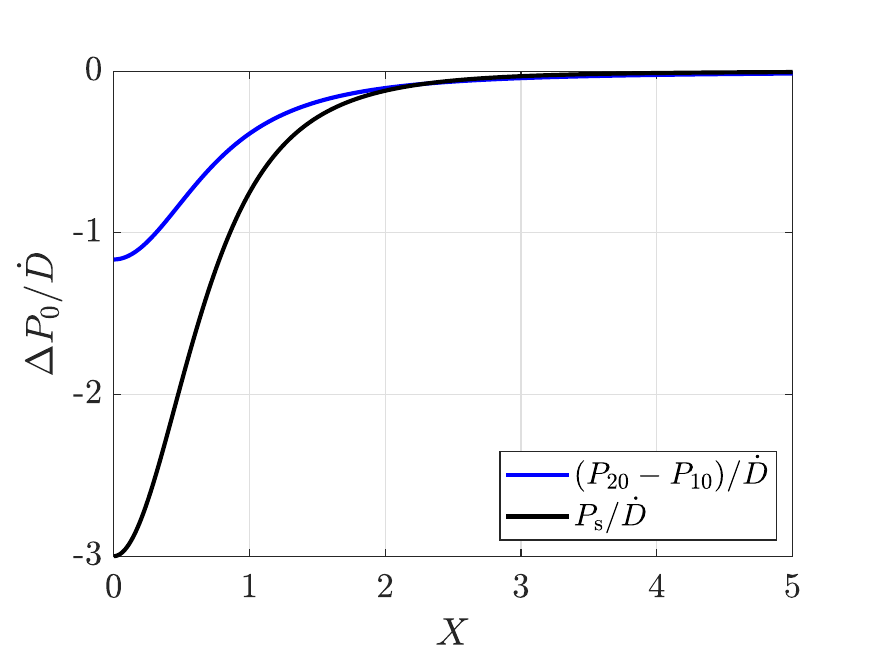}
         \caption{}
     \end{subfigure}
        \caption{a) Zeroth-order excess pressure fields $P_{i0}$, normalized by the cylinder's vertical velocity $\dot{D}$, as functions of horizontal coordinate $X$, as evaluated from Eqs.~(\ref{eq:26})~and~(\ref{eq:27}) with $D = 1$, $M = 1.5$, and $H_{\textrm{b}} = 15$. b) Zeroth-order excess pressure jump $\Delta P_0 = P_{20}-P_{10}$ as a function of horizontal coordinate $X$, as obtained from panel (a). For comparison, we show as well the rigid-case excess pressure $P_{\textrm{s}}=-3\dot{D}/(D+X^2)^2$~\cite{jeffrey1981slow}.}
            \label{fig:2}
\end{figure}
In Fig.~\ref{fig:2}, we plot the zeroth-order excess pressure fields. For comparison, we also show the rigid-case excess pressure $P_{\textrm{s}}=-3\dot{D}/(D+X^2)^2$~\cite{jeffrey1981slow}. As we can see, the pressure fields have opposite signs in the two layers. Moreover, the pressure in the top layer is reduced in the case of an undeformable fluid interface, as compared to the no-slip rigid case. This is due to the fact that horizontal motion at the fluid interface is possible in the former case, which reduces the velocity gradients and stresses.

\begin{figure*}[h]
     \centering
     \begin{subfigure}{0.49\textwidth}
         \centering 
         \includegraphics[width=7cm]{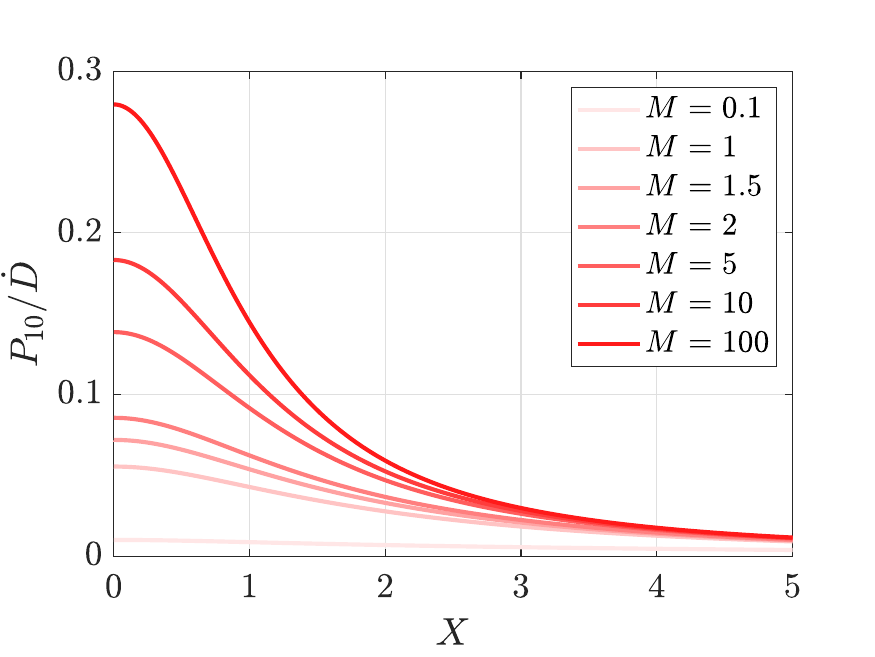}
         \caption{}
         \label{fig:3a}
     \end{subfigure}   
     \begin{subfigure}{0.49\textwidth}
         \centering
         \includegraphics[width=7cm]{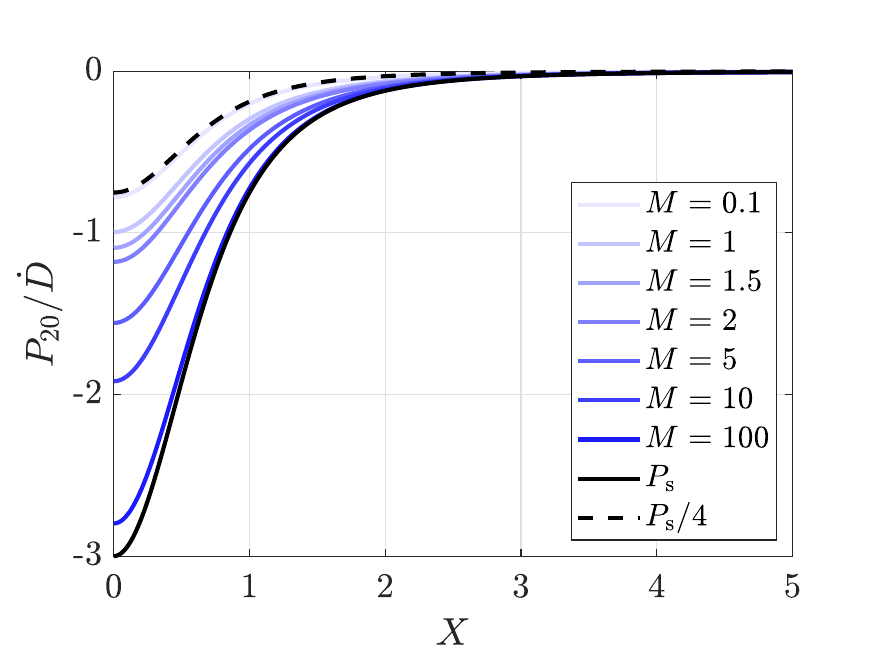}
         \caption{}
         \label{fig:3b}
     \end{subfigure}
        \caption{Zeroth-order excess pressure fields, $P_{10}$ (a) and $P_{20}$ (b), normalized by the cylinder's vertical velocity $\dot{D}$, as functions of horizontal coordinate $X$, as evaluated from Eqs.~(\ref{eq:26})~and~(\ref{eq:27}) with $D = 1$, $H_{\textrm{b}} = 15$ and various $M$ as indicated in legend. For comparison, we show as well the no-slip rigid-case excess pressure $P_{\textrm{s}}=-3\dot{D}/(D+X^2)^2$~\cite{jeffrey1981slow}, and its analogue for a full-slip rigid substrate, that is $P_{\textrm{s}}/4$.}
     \label{fig:3}   
\end{figure*}

Let us now investigate the role of the viscosity ratio $M$. The results are shown in Fig.~\ref{fig:3}. Increasing $M$, \textit{i.e.} increasing the viscosity of the bottom liquid layer as compared to the top one, increases the pressure in both layers. Intuitively, increasing the viscosity ratio makes it harder to generate a flow within the bottom layer, which gets closer to a rigid wall. This is supported by the curves in Fig.~\ref{fig:3}(b) and by Eq.~(\ref{eq:27}), where, at high values of $M$, $P_{20}$ saturates to $P_\textrm{s}$. An interesting point to note in Fig.~\ref{fig:3}(a) and Eq.~(\ref{eq:26}) is that the excess pressure in the bottom layer increases with $M$ as well, but saturates to $9\dot{D}/[2H_{\textrm{b}}(D+X^2)]$ at large $M$, which is dependant on the bottom layer thickness $H_{\textrm{b}}$. On the other hand, if $M$ is decreased towards zero, Eq.~(\ref{eq:26}) predicts that the excess pressure in the bottom layer completely vanishes. Besides, if $M$ is decreased towards zero, Eq.~(\ref{eq:27}) predicts that the excess pressure in the top layer saturates to a quarter of the no-slip rigid-wall value, which is the resulted expected for an effective full-slip interface. The pressure in the top layer is thus bounded at both extremes in $M$. 

\begin{figure*}[h]
     \centering
     \begin{subfigure}{0.49\textwidth}
         \centering
         \includegraphics[width=7cm]{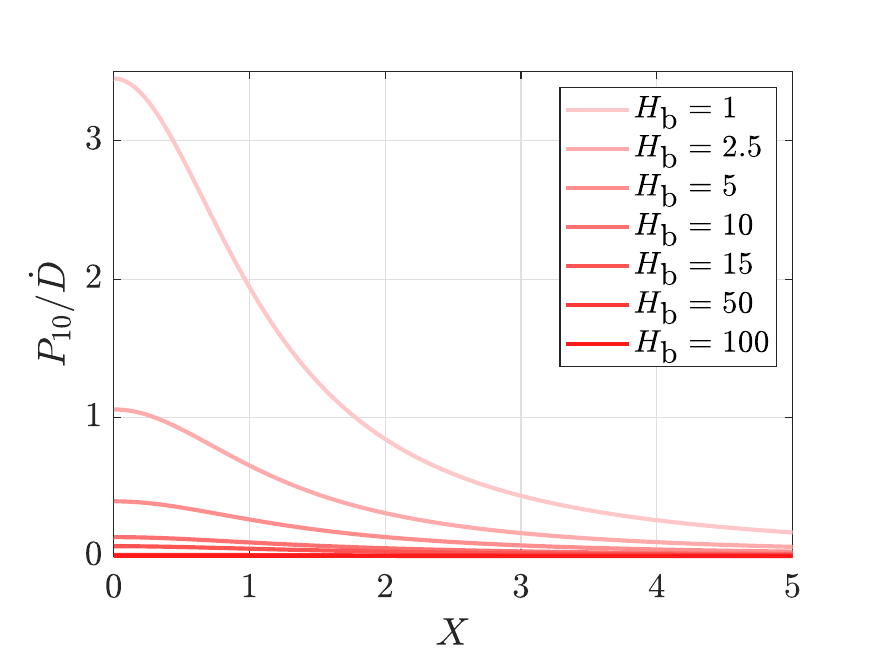}
         \caption{}
         \label{fig:4a} 
     \end{subfigure}
     \begin{subfigure}{0.49\textwidth}
         \centering
        \includegraphics[width=7cm]{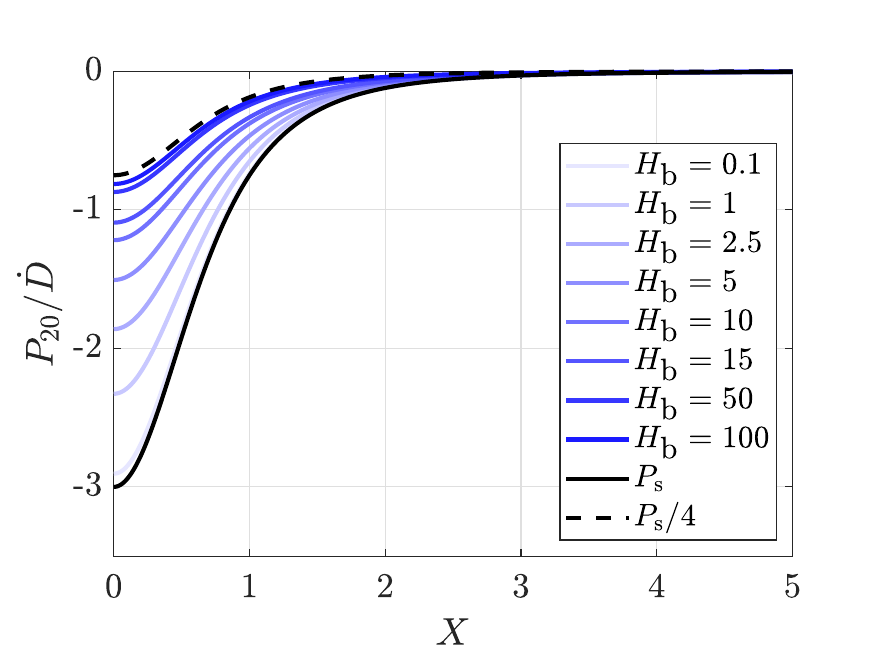}
         \caption{}
         \label{fig:4b} 
     \end{subfigure}
     \caption{Zeroth-order excess pressure fields, $P_{10}$ (a) and $P_{20}$ (b), normalized by the cylinder's vertical velocity $\dot{D}$, as functions of horizontal coordinate $X$, as evaluated from Eqs.~(\ref{eq:26})~and~(\ref{eq:27}) with $D = 1$, $M = 1.5$ and various $H_{\textrm{b}}$ as indicated in legend. For comparison, we show as well the no-slip rigid-case excess pressure $P_{\textrm{s}}=-3\dot{D}/(D+X^2)^2$~\cite{jeffrey1981slow}, and its analogue for a full-slip rigid substrate, that is $P_{\textrm{s}}/4$.}
        \label{fig:4} 
\end{figure*}

The other important parameter to scan and study is the dimensionless bottom-layer thickness $H_\textrm{b}$. Results are shown in Fig.~\ref{fig:4}. Increasing $H_\textrm{b}$ reduces the excess pressure fields in both layers. The two limiting behaviours for $P_{20}$ are the same as when varying $M$, as expected from Eqs.~(\ref{eq:26})~and~(\ref{eq:27}) where it can be seen that the parameter $M/H_\textrm{b}$ is the relevant one. 

\subsection{Interface deflection}
\begin{figure*}[h]
\includegraphics[width=11cm]{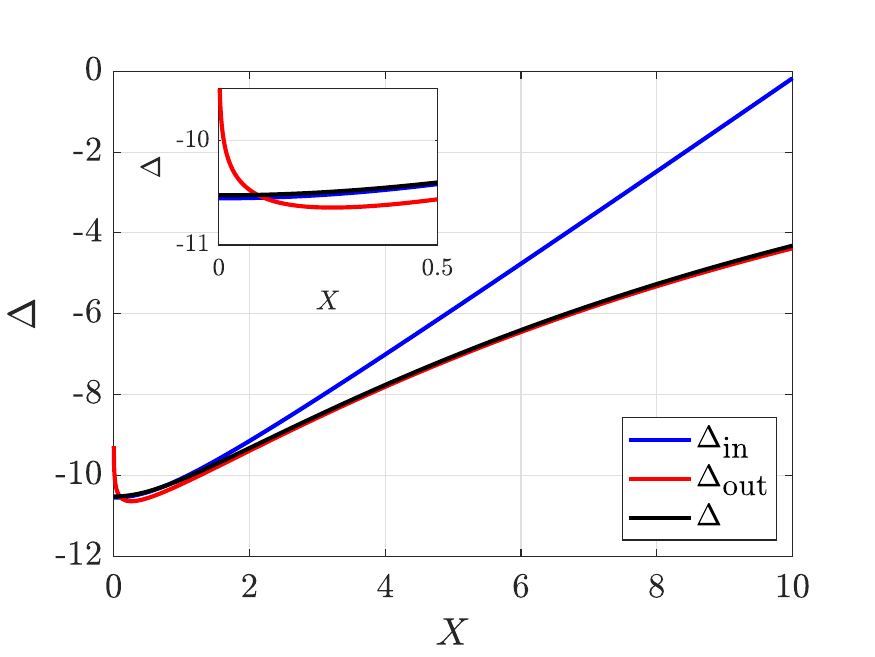}
    \caption{First-order interface deflection $\kappa\Delta$, normalized by dimensionless compliance $\kappa$,  as a function of horizontal coordinate $X$ (black line), as calculated from Eq.~(\ref{gensol}), for  $M = 1.5$, $H_\textrm{b} = 15$, $\textrm{Bo} = 0.01$, $D = 1$ and $\dot{D} = -1$ (\textit{i.e.} cylinder approaching the interface). For comparison, the matched inner (blue) and outer (red) solutions, given by Eqs.~(\ref{eq:innerdelta}) and~(\ref{eq:outerdelta}) respectively, are shown. The inset shows a zoom of the small-$X$ region.}
    \label{fig:interface} 
\end{figure*}

The first-order interface deflection is numerically evaluated from Eq.~(\ref{gensol}) and plotted in Fig.~\ref{fig:interface}, along with the matched inner and outer solutions, given by Eqs.~(\ref{eq:innerdelta}) and~(\ref{eq:outerdelta}). There is a good agreement between the outer and general solutions, except in close proximity to the origin where the outer solution diverges logarithmically. In contrast, while unbounded in the far field, the inner solution agrees well with the general one near the origin.  

\begin{figure*}[h]
     \centering
     \begin{subfigure}{0.49\textwidth}
         \centering
         \includegraphics[width=7cm]{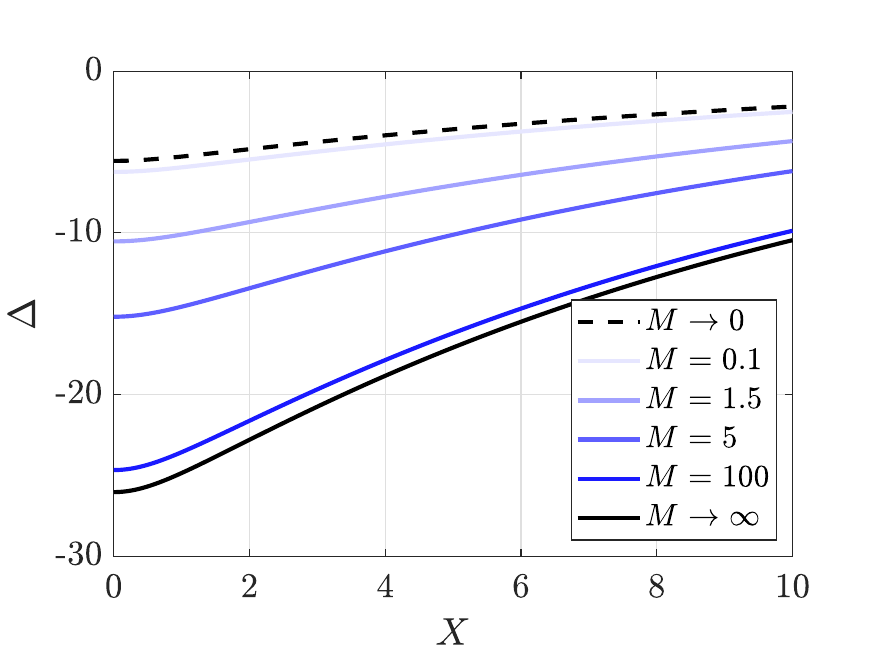}
         \caption{}
         \label{fig:6a} 
     \end{subfigure}
     \begin{subfigure}{0.49\textwidth}
         \centering
         \includegraphics[width=7cm]{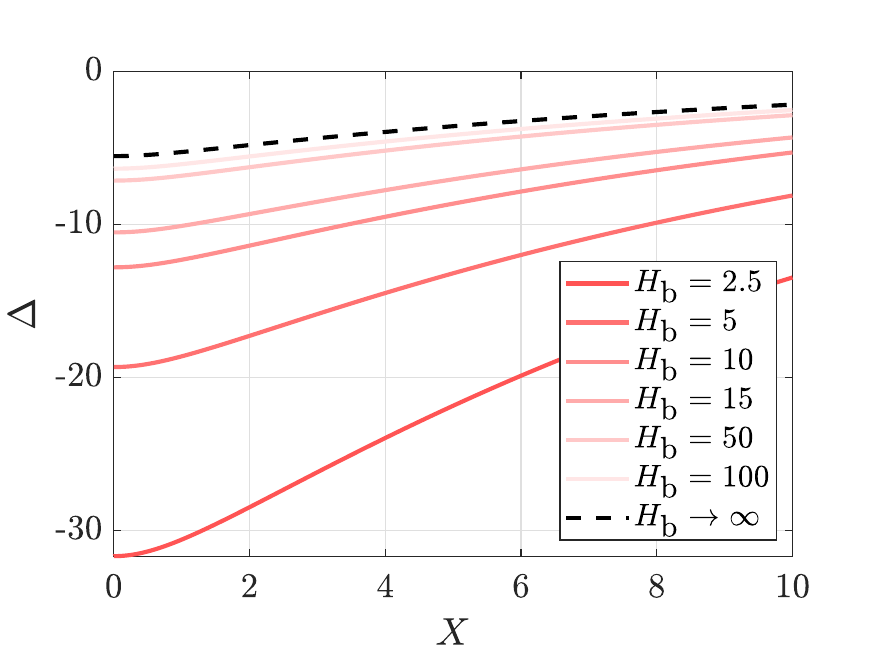}
         \caption{}
         \label{fig:6b} 
     \end{subfigure}
        \caption{a) First-order interface deflection $\kappa\Delta$, normalized by dimensionless compliance $\kappa$, as a function of horizontal coordinate $X$, as calculated from Eq.~(\ref{gensol}), for $H_\textrm{b} = 15$, $\textrm{Bo} = 0.01$, $D = 1$,  $\dot{D} = -1$ (\textit{i.e.} cylinder approaching the interface), and various $M$ as indicated. b) Same as previous panel for $M=1.5$ and various $H_\textrm{b}$ as indicated.}
        \label{fig:6} 
\end{figure*}

Let us now investigate the effects of viscosity ratio $M$ and dimensionless bottom-layer thickness $H_\textrm{b}$ on the interface deflection. The results are shown in Fig.~\ref{fig:6}. We observe that the interface deflection increases with increasing $M$ or decreasing $H_\textrm{b}$. As discussed in the previous section, for vanishing $M$ or infinite $H_\textrm{b}$, the zeroth-order top-layer pressure $P_{20}$ reaches $P_{\textrm{s}}/4$, while the zeroth-order bottom-layer pressure $P_{10}$
vanishes, which leads to the minimal deflection profile. In contrast, as $M$ goes to infinity, $P_{20}$ reaches $P_{\textrm{s}}$, while $P_{10}$ increases to a function depending upon $H_{\textrm{b}}$. Thus, the deflection saturates to a $H_{\textrm{b}}$-dependent profile. However, we stress that decreasing $H_{\textrm{b}}$ increases the deflection without any limit, as the zeroth-order pressure in the bottom layer does not have an upper bound in this case. In reality, such diverging pressure and thus interface deflection would require the consideration of higher-order, non-linear effects in dimensionless compliance.   

\subsection{First-order pressure}
\begin{figure*}[h]
\includegraphics[width=11cm]{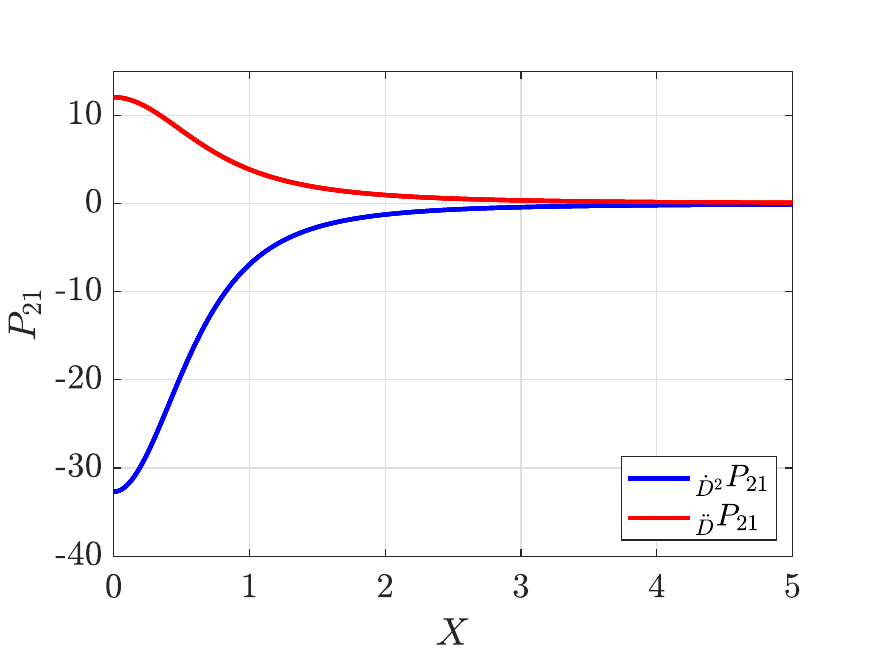}
    \caption{Dynamic adhesive-like (blue) and inertial-like (red) contributions of the first-order pressure correction $\kappa P_{21}$ in the top layer, normalized by the dimensionless compliance $\kappa$, as a function of the horizontal coordinate $X$, obtained from numerical integration of Eqs.~(\ref{eq:thinfilm1}) and~(\ref{eq:thinfilm2}), for  $M = 1.5$, $H_\textrm{b} = 15$, $\textrm{Bo} = 0.01, D = 1, \dot{D} = -1$ and $\ddot{D} = 1$.}
    \label{fig:p21} 
\end{figure*}

\begin{figure*}[h]
     \centering
     \begin{subfigure}{0.49\textwidth}
         \centering
         \includegraphics[width=7cm]{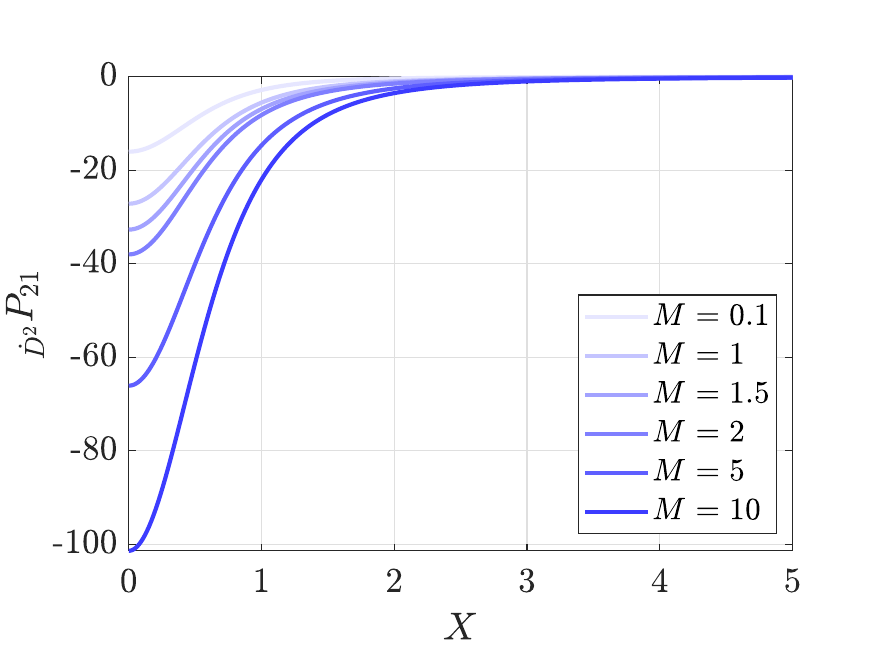}
         \caption{}
         \label{fig:8a} 
     \end{subfigure}
     \begin{subfigure}{0.49\textwidth}
         \centering
         \includegraphics[width=7cm]{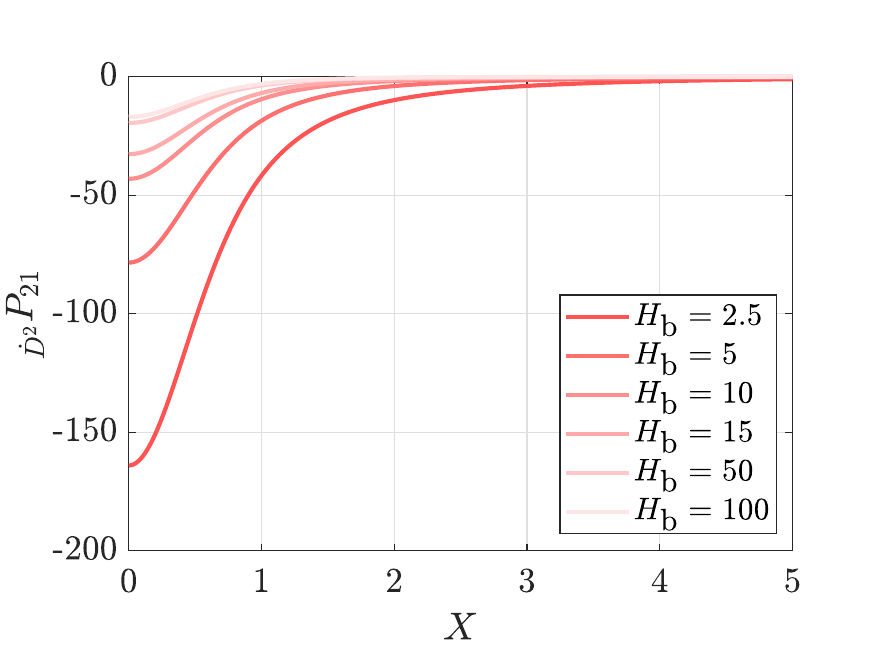}
         \caption{}
         \label{fig:8b} 
     \end{subfigure}
        \caption{a) Dynamic adhesive-like contribution $\kappa{}_{\dot{D}^2} P_{21}$ of the first-order pressure correction in the top layer, normalized by the dimensionless compliance $\kappa$, as a function of the horizontal coordinate $X$, obtained from numerical integration of Eqs.~(\ref{eq:thinfilm1}) and~(\ref{eq:thinfilm2}), for $H_\textrm{b} = 15$, $\textrm{Bo} = 0.01, D = 1, \dot{D} = -1$ and various $M$ as indicated. b) Same as previous panel for $M=1.5$ and various $H_\textrm{b}$ as indicated.}
        \label{fig:8} 
\end{figure*}

\begin{figure*}[h]
     \centering
     \begin{subfigure}{0.49\textwidth}
         \centering
         \includegraphics[width=7cm]{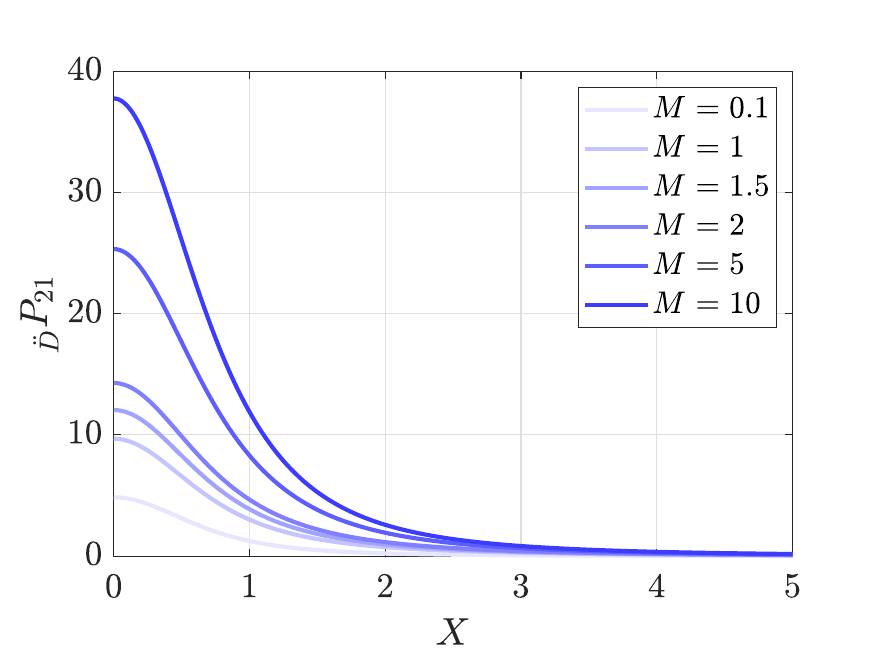}
         \caption{}
         \label{fig:9a} 
     \end{subfigure}
     \begin{subfigure}{0.49\textwidth}
         \centering
         \includegraphics[width=7cm]{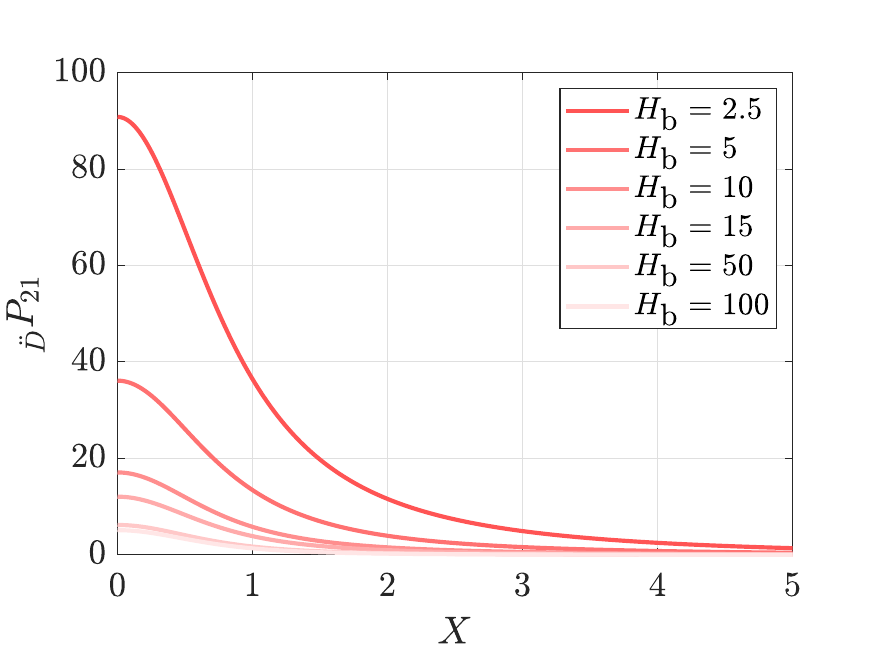}
         \caption{}
         \label{fig:9b} 
     \end{subfigure}
        \caption{a) Inertial-like contribution $\kappa{}_{\ddot{D}} P_{21}$ of the first-order pressure correction in the top layer, normalized by the dimensionless compliance $\kappa$, as a function of the horizontal coordinate $X$, obtained from numerical integration of Eqs.~(\ref{eq:thinfilm1}) and~(\ref{eq:thinfilm2}), for $H_\textrm{b} = 15$, $\textrm{Bo} = 0.01, D = 1, \dot{D} = -1$, $\ddot{D} = 1$ and various $M$ as indicated. b) Same as previous panel for $M=1.5$ and various $H_\textrm{b}$ as indicated.}
        \label{fig:9} 
\end{figure*}

Numerically integrating Eqs.~(\ref{eq:thinfilm1}) and~(\ref{eq:thinfilm2}) at first order in $\kappa$ allows us to find the first-order pressure correction. Note that we only consider the first-order top-layer pressure $\kappa P_{21}$ for two reasons. First, this contribution is the only one required to eventually compute the first-order force exerted on the cylinder (see next section). Secondly, the first-order bottom-layer pressure decays slowly in $X$, and seems to depend on the size of the numerical window, indicating the potential need for a far-field regularization. Besides, we decompose $P_{21}$ into two contributions: i) a dynamic adhesion-like term ${}_{\dot{D}^2} P_{21}$, depending on the square of the vertical velocity of the cylinder, which tends to attract the moving object towards the deformable interface \cite{kaveh2014hydrodynamic,salez2015elastohydrodynamics,wang2015out,bertin2022soft}; and ii) an inertial-like term ${}_{\ddot{D}} P_{21}$, depending on the vertical acceleration of the cylinder, which is present here as a consequence of volume conservation \cite{salez2015elastohydrodynamics,bertin2022soft}, even though the governing equations are free of inertia. The results are shown in Figs.~\ref{fig:p21},~\ref{fig:8} and~\ref{fig:9}. We see that both pressure contributions are maximal at the centre ($X=0$) and vanish quickly above $X\sim1$. Besides, changing the viscosity ratio $M$ and the dimensionless bottom-layer thickness $H_{\textrm{b}}$ have the same effects as for the zeroth-order case. 

\subsection{Capillary-lubrication force}
Since viscous stresses are negligible as compared to the excess pressure field within the lubrication framework, and since the excess pressure field obtained above typically vanishes beyond $X\sim1$ in the top layer, the normal capillary-lubrication force per unit length felt by the cylinder can be simply evaluated by integrating the excess pressure field in the top layer along the horizontal coordinate. Putting back dimensions, the force per unit length thus reads at first order in compliance:
\begin{align} 
 F &= \int_{-\infty}^{+\infty}\textrm{d}x\, p_2\\  &\simeq -\eta_2\dot{d}\left(\frac{a}{d}\right)^{3/2}\phi_0(M,H_{\textrm{b}},D)- \frac{\eta_2^2\dot{d}^2}{\sigma}\left(\frac{a}{d}\right)^{7/2}{}_{\dot{D}^2}\phi_1(M,H_{\textrm{b}},\textrm{Bo},D)+\frac{\eta_2^2\ddot{d}a}{\sigma}\left(\frac{a}{d}\right)^{5/2}{}_{\ddot{D}}\phi_1(M,H_{\textrm{b}},\textrm{Bo},D),    \label{eq:41}   
\end{align}
where $\phi_0$, ${}_{\dot{D}^2}\phi_1$ and ${}_{\ddot{D}}\phi_1$ are auxiliary functions depending on the parameters of the problem, $M$, $H_{\textrm{b}}$, as well as $\textrm{Bo}$ and, importantly, potentially having extra dependencies in $d$ through $D$.
\begin{figure}[h]
\includegraphics[width=11cm]{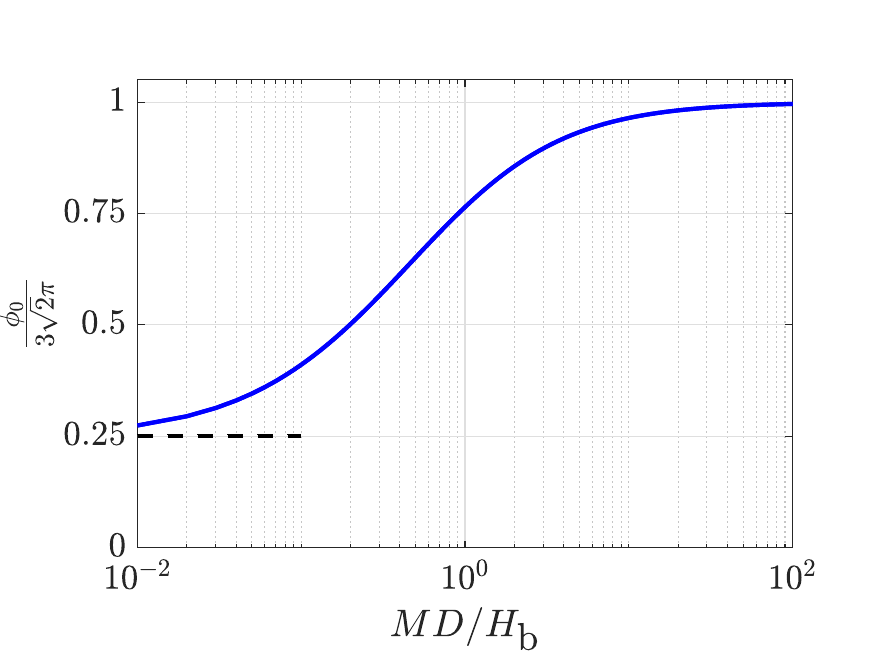}
\caption{Zeroth-order auxiliary function $\phi_0$ of the normal force (see Eq.~(\ref{eq:41})), normalized by the corresponding value $3\sqrt{2}\pi$ of the no-slip rigid case~\cite{jeffrey1981slow}, as a function of the single rescaled variable $MD/H_\textrm{b}$. The dashed line shows a constant value of 1/4.}
\label{fig:phi0variation}
\end{figure}

Let us first study the zeroth-order contribution to the force, through the auxiliary function $\phi_0$. Using Eq.~(\ref{eq:27}), one can evaluate $\phi_0$ and show that it depends only on the variable $MD/H_{\textrm{b}}$. The function is shown in Fig.~\ref{fig:phi0variation}. It is always positive, indicating a Stokes-like drag effect. At infinite $MD/H_{\textrm{b}}$ one recovers the no-slip rigid case~\cite{jeffrey1981slow}, and the scaling of the force with $d$ is thus $\sim d^{-3/2}$. At vanishing $MD/H_{\textrm{b}}$, $\phi_0$ saturates to a quarter of the no-slip rigid value, which corresponds to the case of a full-slip rigid wall, with pressure $P_{\textrm{s}}/4$ as discussed above, and the scaling is once again $\sim d^{-3/2}$. In between these limits, we observe a smooth crossover and there is thus no clear power law in $d$. 
 \begin{figure*}[h]
     \centering
     \begin{subfigure}{0.49\textwidth}
         \centering
         \includegraphics[width=7cm]{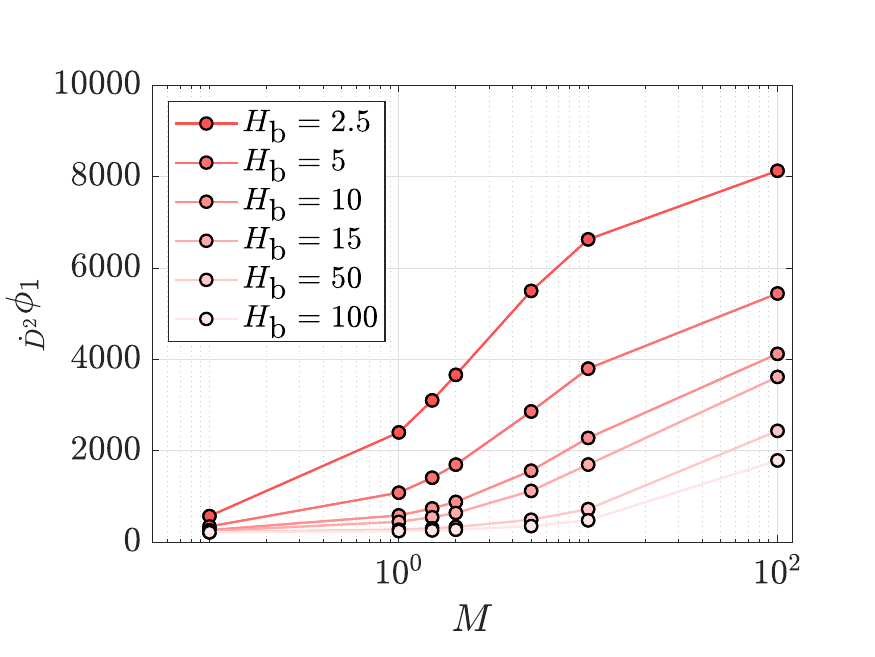}
         \caption{}
         \label{fig:phi1ddotsqvar} 
     \end{subfigure}
     \begin{subfigure}{0.49\textwidth}
         \centering
         \includegraphics[width=7cm]{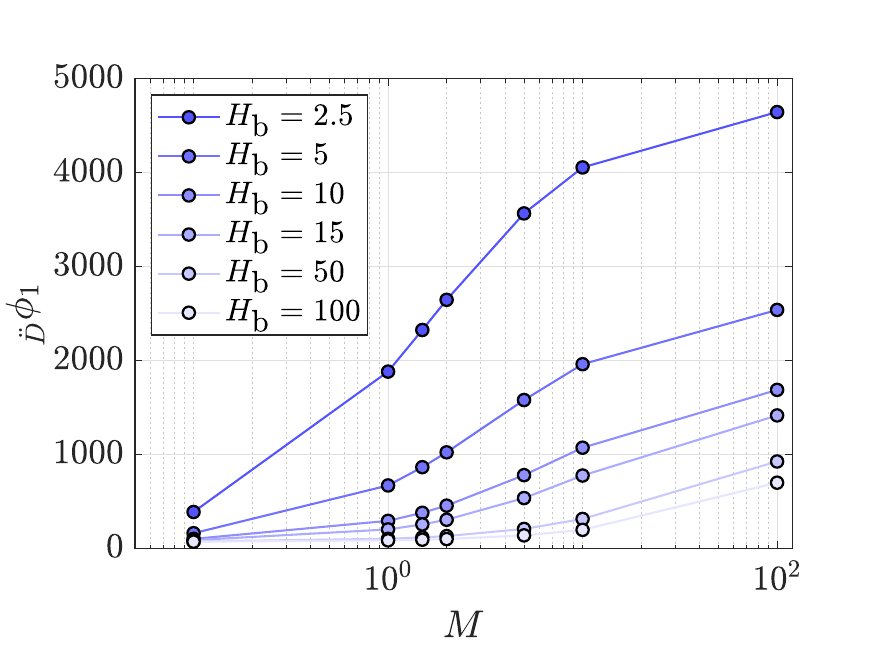}
         \caption{}
         \label{fig:phi1ddotdotvar} 
     \end{subfigure}
    \caption{First-order auxiliary functions, ${}_{\dot{D}^2}\phi_1$ (a) and ${}_{\ddot{D}}\phi_1$ (b), of the normal force (see Eq.~(\ref{eq:41})), as functions of the viscosity ratio $M$, as obtained from numerical integration of the first-order excess pressure $P_{21}$ in the top layer, for $\textrm{Bo}=0.01$, $D=1$, and for various values of the dimensionless bottom-layer thickness $H_{\textrm{b}}$, as indicated. The lines are guides for the eye.}
    \label{fig:11} 
\end{figure*}

Finally, we study the first-order contributions to the force, through the two auxiliary functions ${}_{\dot{D}^2}\phi_1$ and ${}_{\ddot{D}}\phi_1$. In these cases, there does not seem to be a simple combination of the parameters $M$, $H_{\textrm{b}}$, $\textrm{Bo}$ and variable $D$, that controls the auxiliary functions. These are numerically evaluated by integration of $P_{21}$, and plotted in Fig.~\ref{fig:11} for various parameters. The two auxiliary functions are always positive and grow with increasing $M$ or decreasing $H_{\textrm{b}}$. They seem to saturate at either vanishing $M$ or infinite $H_{\textrm{b}}$. At infinite $M$ there might be a saturation as well. However, the functions seem unbounded when decreasing $H_{\textrm{b}}$. As a final remark, the important increase observed for the auxiliary functions in some parametric ranges enforces stringent conditions on the dimensionless compliance $\kappa$ so as to be in line with the perturbative approach.   

\section{Conclusion}
We have theoretically and numerically studied the capillary-lubrication force felt by an immersed infinite cylinder when approaching towards a fluid interface. To do so, we invoked lubrication theory and performed a perturbation analysis in dimensionless compliance, where the latter was expressed in terms of a capillary number and a lubrication parameter. The zeroth-order excess pressure fields in the two liquid layers were calculated analytically, and exhibit saturations towards solid-like behaviours when the bottom layer becomes extremely viscous or thin. We then numerically computed the interface deflection at first order in compliance, and analysed its near-field and far-field behaviours thanks to  asymptotic matching, at small Bond number. An interesting outcome is the estimate of the central deformation of the fluid interface. Then, the excess pressure fields at first order in compliance were numerically computed. We identified two main dynamic contributions: i) a velocity-dependent adhesive-like one, and ii) an acceleration-dependent inertial-like one. In all the steps, we investigated the effects of two key dimensionless parameters : the viscosity and thickness ratios. Our results might find applications in confined colloidal systems.

\begin{acknowledgments}
The authors thank Vincent Bertin, Andreas Carlson, and Christian Pedersen for interesting discussions. They acknowledge financial support from the European Union through the European Research Council under EMetBrown (ERC-CoG-101039103) grant. Views and opinions expressed are however those of the authors only and do not necessarily reflect those of the European Union or the European Research Council. Neither the European Union nor the granting authority can be held responsible for them. The authors also acknowledge financial support from the Agence Nationale de la Recherche under EMetBrown (ANR-21-ERCC-0010-01), Softer (ANR-21-CE06-0029), and Fricolas (ANR-21-CE06-0039) grants. Finally, they thank the Soft Matter Collaborative Research Unit, Frontier Research Center for Advanced Material and Life Science, Faculty of Advanced Life Science at Hokkaido University, Sapporo, Japan.  
\end{acknowledgments}
\bibliography{mainbib}
\end{document}